\documentclass[aps,prc,letterpaper,11pt,twoside,tightenlines,nofootinbib,showpacs,preprint]{revtex4}
\usepackage{graphicx}
\usepackage[sort&compress]{natbib}
\usepackage{subfigure}
\usepackage{amsmath}
\usepackage{amsfonts}
\usepackage{cancel}
\usepackage{lmodern,dsfont}
\usepackage{amssymb}
\begin{document}
\arraycolsep1.5pt
\newcommand{\Ima}{\textrm{Im}}
\newcommand{\Rea}{\textrm{Re}}
\newcommand{\mev}{\textrm{ MeV}}
\newcommand{\be}{\begin{equation}}
\newcommand{\ee}{\end{equation}}
\newcommand{\bea}{\begin{eqnarray}}
\newcommand{\eea}{\end{eqnarray}}
\newcommand{\bef}{\begin{figure}}
\newcommand{\eef}{\end{figure}}
\newcommand{\bce}{\begin{center}}
\newcommand{\ece}{\end{center}}
\newcommand{\ba}{\begin{eqnarray}}
\newcommand{\ea}{\end{eqnarray}}
\newcommand{\gev}{\textrm{ GeV}}
\newcommand{\nn}{{\nonumber}}
\newcommand{\dtres}{d^{\hspace{0.1mm} 3}\hspace{-0.5mm}}
\newcommand{\rts}{ \sqrt s}
\newcommand{\non}{\nonumber \\[2mm]}

\title{ $\bar{K}NN$  Absorption within the Framework of the Fixed Center Approximation to Faddeev equations}

\author{M. Bayar$^1$,$^2$ and E. Oset$^1$}
\affiliation{$^1$Instituto de F{\'\i}sica Corpuscular (centro mixto CSIC-UV)\\
Institutos de Investigaci\'on de Paterna, Aptdo. 22085, 46071, Valencia, Spain,\\ $^2$Department of Physics, Kocaeli University, 41380 Izmit, Turkey}

\begin{abstract}

We present a method to evaluate the  $\bar{K}$ absorption width in the bound $\bar{K}NN$ system. Most calculations of this system  ignore this channel and only consider the
$\bar{K}N \rightarrow \pi \Sigma$ conversion. Other works make a qualitative calculation using perturbative methods. Since the  $ \Lambda(1405) $ resonance is playing a role in the process, the same resonance is changed by the presence of the absorption channels and we find that a full nonperturbative calculation is called for, which we present here. We employ the Fixed Center Approximation to Faddeev equations to account for $\bar{K}$ rescattering on the $ (NN) $ cluster and we find that the width of the states found previously for $ S=0 $ and $ S=1 $ increases by about 30 MeV due to the $\bar{K}NN$ absorption, to a total width of about 80 MeV.

\end{abstract}


\vspace{1cm}

\date{\today}

\maketitle

\section{Introduction}
 \label{sec:intro}
 
 The interaction of antikaon ($\bar{K}  $) with nucleon ($ N $) has drawn attention for decades   \cite{Kaiser:1995eg,angels,Oller:2000fj,Lutz:2001yb,Oset:2001cn,Hyodo:2002pk,Jido:2003cb,Borasoy:2005ie,Oller:2006jw,Borasoy:2006sr,Ikeda:2012au} since this interaction provides essential elements to understand the strangeness in nuclear hadron physics. The $\bar{K} N $ interaction is strongly attractive in the isospin $ I=0 $  and also the  $ \pi \Sigma $ interaction is attractive at low energies. The $\bar{K} N $ and $ \pi \Sigma $ are the most important channels for the $ \Lambda(1405) $: these two channels dynamically generate the quasi bound $ \Lambda(1405) $.

 To understand the many body kaonic nuclear system, the $\bar{K}NN $ is the simplest system. The investigation of this system started in the 60's \cite{nogami}. 
 More recently, using the variational approach, a $\bar{K}NN $ state was found bound by 48 MeV with width of 60 MeV \cite{Yamazaki:2002uh}. 
   In \cite{Dote:2008in}, the $ K^- pp $ three body system was investigated using the variational approach taking the results of $\bar{K}N  $ interactions from a chiral $ SU(3) $ model, and they get a much less bound $ K^- pp $ state with, B=19 MeV and $ \Gamma $=40-70 MeV. In a further paper, the same group quantifies the uncertainties and evaluates the extra width coming from $\bar{K}$ absorption from the pair of nucleons, with the results of  B=20-40 MeV and  $ \Gamma $ as large as 100 MeV with admitted large uncertainties \cite{Dote:2008hw} . Coupled channel three body calculations of the quasi bound $\bar{K}NN $ system were investigated in \cite{Shevchenko:2006xy,Shevchenko:2007zz}  using Faddeev equations,  and their results for the binding is 50-70 MeV and the width around 100 MeV. The work of \cite{Ikeda:2007nz} studied three body resonances in the $\bar{K}NN $ system within a framework of the $\bar{K}NN \rightarrow \pi  Y N$ coupled channel Faddeev equation. In this work they found the binding energy 79 MeV with a width of 74 MeV. However, in more recent papers \cite{Ikeda:2008ub, Ikeda:2010tk}, where the energy dependence of the potential is taken from chiral dynamics, the same authors find much smaller binding energies,  below 20 MeV.  We also calculated the  $\bar{K}NN $ system \cite{bayarnpa,  Oset:2012gi} for the $ S=0 $ and $ S=1 $ case, using the Fixed Center Approximation (FCA) to the Faddeev equations taking into account $ \pi \Sigma N $ channel explicitly and also including the charge exchange diagrams. We get a binding of 26-35 MeV for $ S=0 $ while around 9 MeV for the $ S=1 $ case. The widths are in both cases around 50 MeV. A very recent  variational calculation \cite{Barnea:2012qa} finds that the $ S=0 $ state is bound by about 16 MeV and the  $ S=1 $ by less than 11 MeV. The widths, without including $\bar{K} $ absorption by a pair of nucleons, are around 40 MeV.

The FCA to the Faddeev equations is an efficient and useful method to investigate many body systems. It relies upon having a pair of particles that interact strongly among themselves and make a cluster that is supposed not to be much disturbed by the interaction of the third particle with it. Then all that is left is the possibility that the third particle interacts with the components of the cluster and this is done non perturbative allowing for multiple scattering of this third particle with any of the components of the cluster. The cluster is then assumed to have the wave function that it would have without the presence of the third particle.  Intuitively, one should expect it to work when the third particle is light compared to those of the cluster, or when it is very bound such that it cannot allow excitations of the cluster. There is substantial work to rely on this model. For instance, the three body $ \bar{K} NK $ scattering amplitude  \cite{Xie:2010ig} was calculated using the FCA to the Faddeev equations and the results of this work are in good agreement with the other theoretical works \cite{Jido:2008kp, MartinezTorres:2008kh}  evaluated using variational and Faddeev approaches, respectively. Besides, in  \cite{Xie:2011uw}, using the same model, the authors give a plausible explanation for the $ \Delta_{\frac{5}{2}^{+}} (2000) $ puzzle. As important as to know the success of the FCA, it is to know the limitation of this procedure, and it was found in  \cite{MartinezTorres:2010ax} that the approximation collapses for the study of resonant states, where there is plenty of phase space for the excitation of intermediate states.

Meson nucleus bound systems are good laboratories to investigate the finite density QCD. The existence of the $ \bar{K} $ bound states in nuclear systems, which are called kaonic nuclei, is theoretically expected because of the strong attractive interactions between the $ \bar{K} $   and the nucleon. Models range from empirical ones with a depth of around 200 MeV at normal nuclear matter density \cite{Friedman:1999rh}, to those imposing chiral symmetry constraints that lead to an attraction of around 40-50 MeV \cite{Koch:1994mj,Lutz:1997wt,Ramos:1999ku,SchaffnerBielich:1999cp,Cieply:2001yg}.
 However there is no  experimental evidence on this system.  
In spite of experimental claims, which have been proved to be unfounded  (see Ref. \cite{Ramos:2008zza} for a review on the issue). The large width of the predicted states compared to the binding could justify why such states are not being found  \cite{Hirenzaki:2000da}. 
The $ K^- pp $ is the prototype of the $\bar{K}  $ nuclei and to observe the kaonic nuclei in experiments, the precise knowledge of the width of this state is important. In this sense, it is important to recall that in all previous works of the $\bar{K}NN $ system the source of the width is only the  $\bar{K} N \rightarrow \pi \Sigma $ conversion channel. The absorption channels $\bar{K}NN \rightarrow \Lambda N, \Sigma N $ are not considered explicitly, although in some case it is estimated perturbatively \cite{Dote:2008hw}.
 Early experiments on $ K^- $-nucleus absorption at rest were done in the 1970s. A precise measurement of the total  two nucleon absorption of stopped $ K^- $ mesons in deuterium  was reported in \cite{Veirs:1970fs} using deuterium bubble chambers. This is the first detailed measurement of two nucleon absorption branching ratio and these results were compared with the predictions of isospin invariance for strong interaction. The ratio of the rates $ R(K^-d \rightarrow \Sigma^- p ) $ and  $ R(K^-d \rightarrow \Sigma^0 n ) $ was in good agreement with the results of isospin invariance for strong interaction. A Similar experiment was done using a helium bubble chamber \cite{Katz:1970ng}.
 
 $\bar{K} $ absorption by two nucleons in nuclei, $\bar{K}NN \rightarrow \Lambda N, \Sigma N , \Sigma^{*} N $, has been considered before in the selfconsistent evaluation of the $\bar{K}$ nucleon optical potential of \cite{Ramos:1999ku,Oset:2000eg}. More recently it has also been evaluated in a different way in \cite{Sekihara:2012wj}. The $\bar{K}$ nucleus optical potential evaluated in \cite{Ramos:1999ku} has been checked versus kaonic atoms in \cite{Hirenzaki:2000da} with good agreement with experiment. The discrepancies seen for the $ ^{4}He $ atom have been revolved recently with very precise measurement in \cite{Okada:2007ky,Bazzi:2009zz} which agree with the early predictions of \cite{Hirenzaki:2000da}. Also the study done in \cite{Baca:2000ic} showed that a best fit to the data could be obtained with a potential that differed from the predicted one of \cite{Ramos:1999ku} at the level of 20 $ \% $. This agreement with data gives us confidence that the input used in \cite{Ramos:1999ku} for the $ \bar{K}NN $ absorption is realistic and we use this input here to evaluate the absorption width of the $\bar{K}NN$ state.

The article is organized as follows. In Section II, the calculation of the three body $\bar{K} N N$ amplitude including the charge exchange mechanisms is summarized using the Faddeev equations under the FCA. In Section III, The explicit derivation of the $ K^- $ absorption by two nucleons is given. The results of the  $ K^- $ absorption both for spin-0 and spin-1 are shown in Section IV.

\section{ Calculation of the $\bar{K} N N$ amplitude}
Here we will summarize the derivation of the $\bar{K} N N$ three body amplitude within the framework of the FCA to the Faddeev equations taking into account the charge exchange contributions.
In order to investigate the three-body amplitude of the $\bar{K} N N$ system, we have two possible spin states, $ S=0 $ and $ S=1 $. Let us first concentrate on the calculation of the $ S=0 $ case which is done in detail in \cite{bayarnpa}. At the end we want total isospin-$ \frac{1}{2} $ for three-body amplitude. 
The wave function for this state is 
\begin{equation}
|K^- p p>=-(\dfrac{1}{\sqrt{3}}|3/2,1/2>+\sqrt{\dfrac{2}{3}}|1/2,1/2>)
\end{equation}
with the basis of $|I_{tot},I_{3,tot}>$. For the total amplitude with total isospin-$ \frac{1}{2} $  we have 
 \begin{equation}
<1/2|T|1/2>=\dfrac{3}{2}(<K^- p p|T|K^- p p> -\dfrac{1}{3}<3/2|T|3/2>).
\label{Eq:tttal}
\end{equation}

\begin{figure}[h]
\centering
\includegraphics[width=0.3\textwidth] {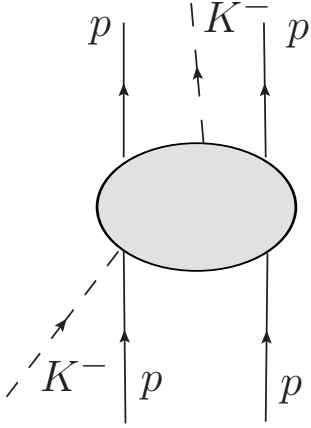}
\caption{Diagrammatic representation of the $K^- (p p) $ interaction. The equivalent diagram where $ K^-$ first interacts with the second proton should be added.}
\label{fig:kmppam}
\end{figure}

\begin{figure}
\centering
\includegraphics[width=0.9\textwidth] {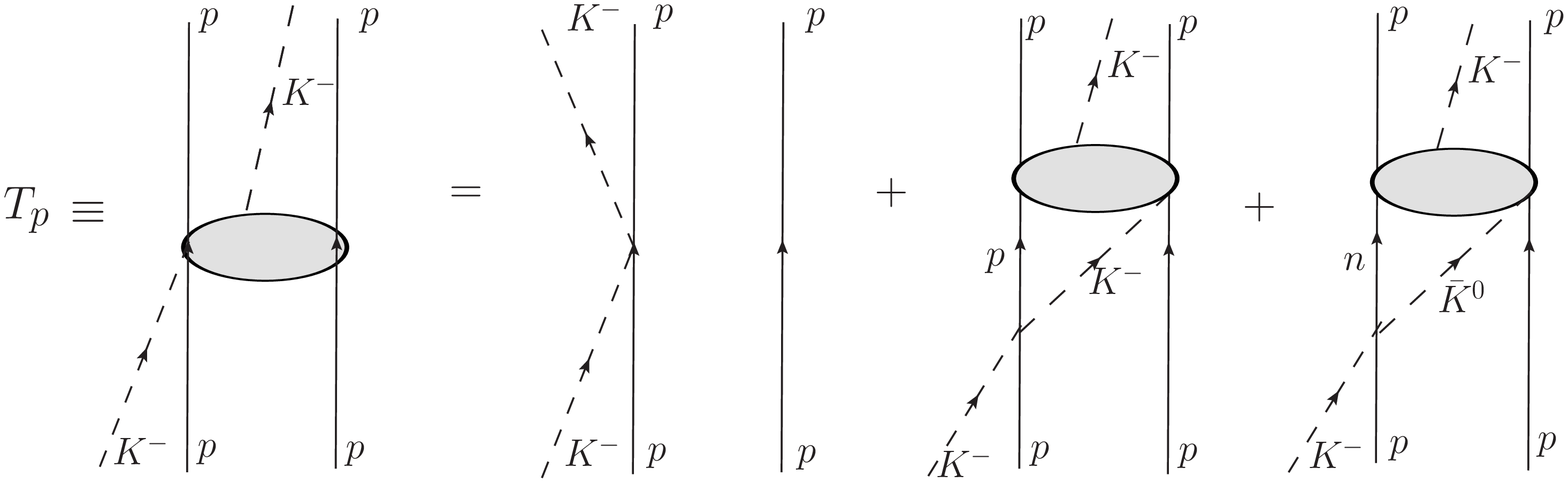}\\
\includegraphics[width=0.5\textwidth] {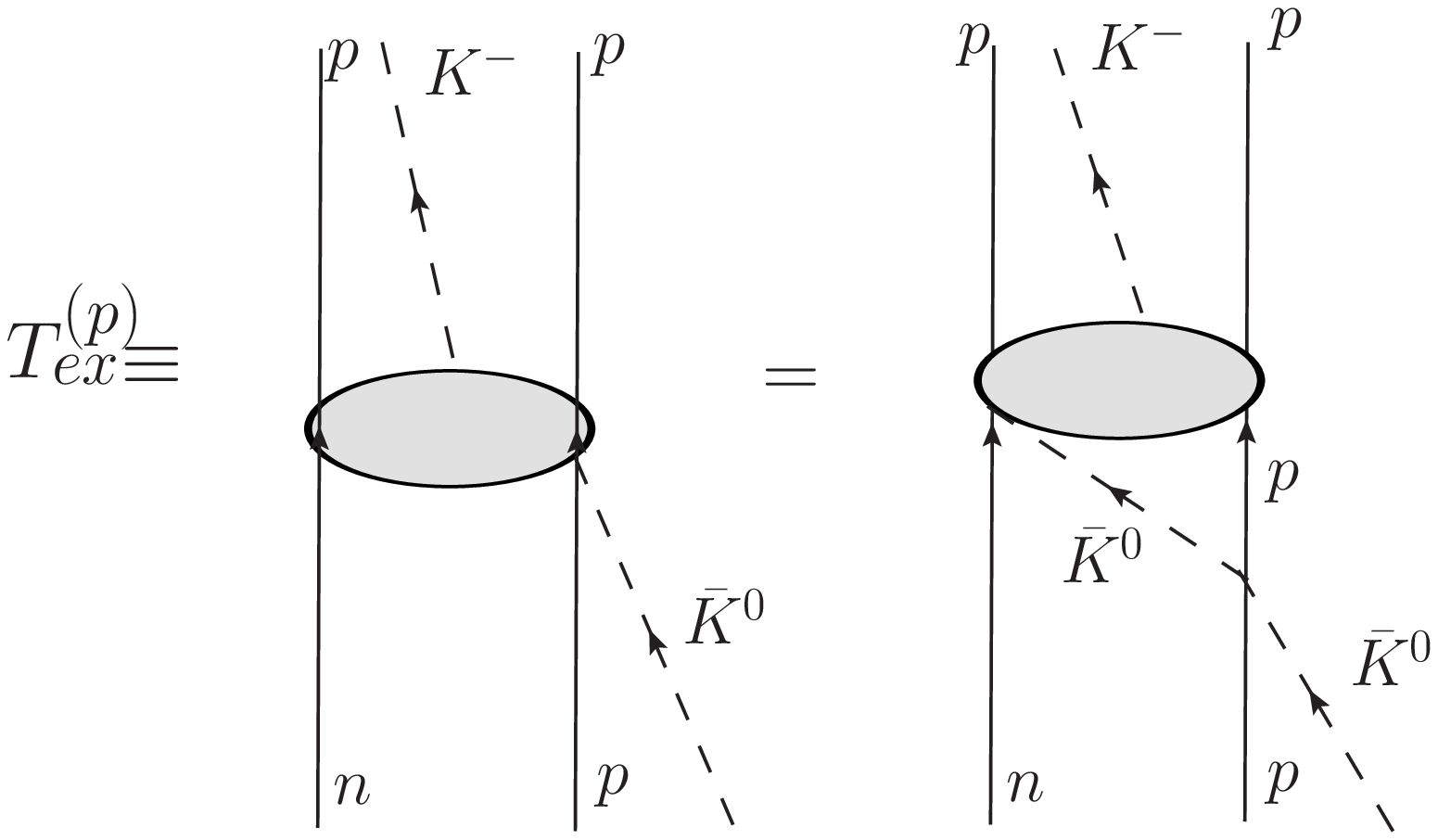}\\
\includegraphics[width=0.9\textwidth] {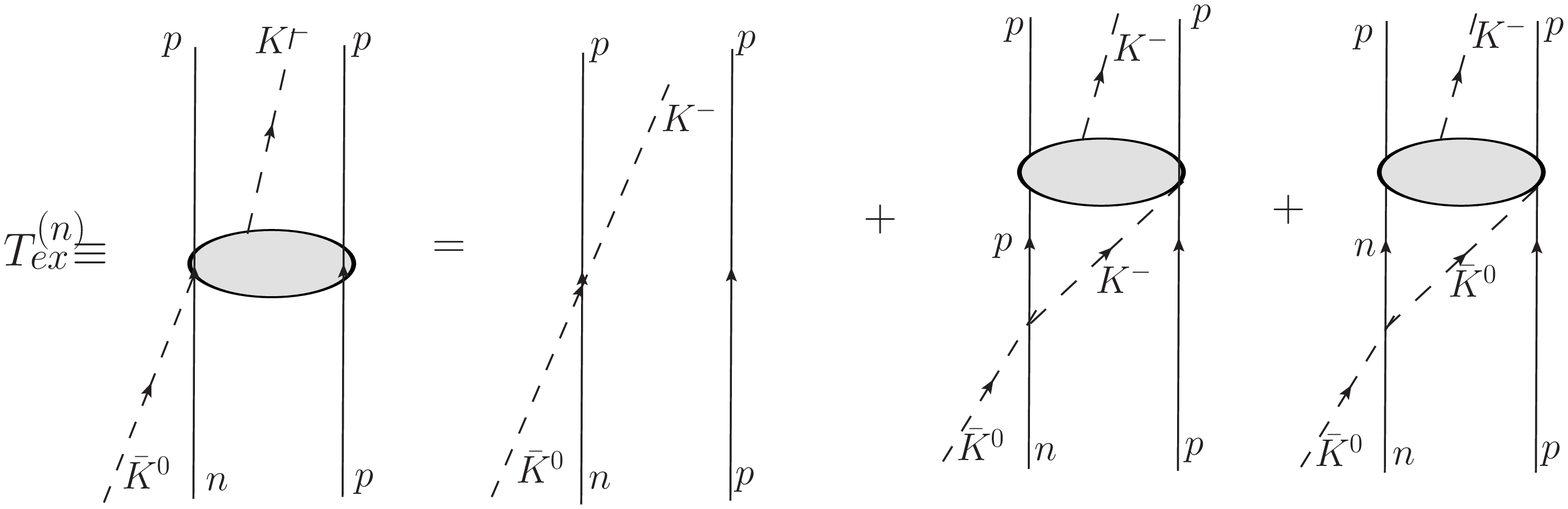}
\caption{Diagrammatic representation of the partition functions for $K^- p p \rightarrow K^- p p $.}
\label{fig:tppdiag}
\end{figure}

First,  we start from the $K^- p p \rightarrow K^- p p$ amplitude where a $ K^- $ interacts with either of the two protons and the diagrammatic representation of  this amplitude is shown in Fig. \ref{fig:kmppam}, where the shaded ellipse includes all multiple scattering of the $ K^- $ where the $ K^- $ interacts first with the first nucleon. We call the result of this diagram $ T_{p} $ and hence the total amplitude including the diagrams where the $ K^- $ interacts first with the second nucleon is $ 2 T_{p} $.
Looking in detail at the ellipse in Fig. \ref{fig:kmppam} we have three partition functions (see Fig. \ref{fig:tppdiag}) that fulfill the following coupled channel equations  
\begin{eqnarray}
T_{p}&=&t_{p}+t_{p}G_0T_{p}+t_{ex}G_0T_{ex}^{(p)}\nonumber\\
T_{ex}^{(p)}&=&t_{0}^{(p)}G_0T_{ex}^{(n)}\nonumber\\
T_{ex}^{(n)}&=&t_{ex}+t_{ex}G_0T_{p}+t_{0}^{(n)}G_0T_{ex}^{(p)}
\label{Eq:partf}
\end{eqnarray} 
where
$t_p=t_{K^- p , K^- p}$, $t_{ex}=t_{K^- p , \bar{K}^0 n}$, $t_{0}^{(p)}=t_{\bar{K}^0 p , \bar{K}^0 p}$, $t_{0}^{(n)}=t_{\bar{K}^0 n , \bar{K}^0 n}$
and $G_0$  \cite{Bayar:2011qj,multirho} is 

\begin{equation}
G_0=\int\frac{d^3q}{(2\pi)^3}F_{NN}(q)\frac{1}{{q^0}^2-\vec{q}\,^2-m_{\bar K}^2+i\epsilon},
\label{Eq:gzero}
\end{equation}
with the factor $ F_{NN}(q) $ standing for the form factor of the bound $ NN $ cluster.
The diagrammatic representation of the partition functions is shown in Fig. \ref{fig:tppdiag}. Calculating the three equations in Eq. (\ref{Eq:partf}) we get $ T_{p} $  as below

\begin{equation}
T_{p}=\frac{t_{p}(1-t_{0}^{(n)}G_0t_{0}^{(p)}G_0)+t_{ex}^2G_0t_{0}^{(p)}G_0}{(1-t_{p}G_0)(1-t_{0}^{(n)}G_0t_{0}^{(p)}G_0)-t_{ex}^2t_{0}^{(p)}G_0^3}.
\label{Eq:Tchare}
\end{equation}

\begin{figure}
\centering
\includegraphics[scale=0.5]{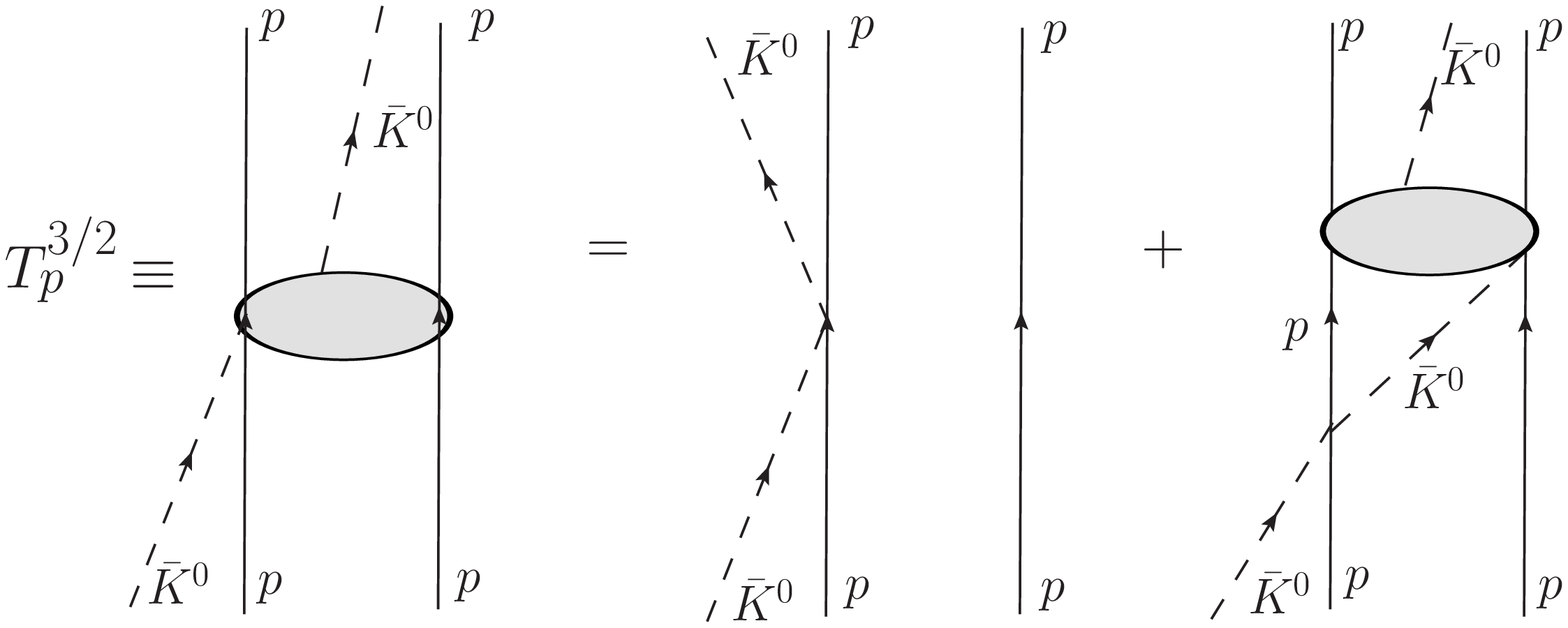}
\caption{Diagrammatic representation of the partition function for I=3/2.}\label{tppI32}
\end{figure}   

Now we need to evaluate the second term of Eq. (\ref{Eq:tttal}). This term does not have charge exchange and its diagrammatic representation is shown in Fig. \ref{tppI32}. Taking into account the equivalent diagram where $ \bar{K}^{0} $ interacts first with the second nucleon we obtain
\begin{equation}
T_p^{(3/2)}=2\dfrac{t_{0}^{(p)}}{1-G_0t_{0}^{(p)}}.
\end{equation}
Using Eq. (\ref{Eq:tttal}), the final result of the amplitude for $ S=0 $ case is 
\begin{eqnarray}
T^{(1/2)}=3T_p-\dfrac{t_{0}^{(p)}}{1-G_0t_{0}^{(p)}}.
\label{Eq:bizim}
\end{eqnarray}
In the case of $ S=1 $, the amplitude of the three-body system is calculated in Ref. \cite{Oset:2012gi} following the steps of \cite{kamalov}  for the evaluation of the $ K^{-}d $ scattering length. The resulting amplitude is 

\begin{equation}
T_{K^- d}=\frac{t_p+t_n+(2t_pt_n-t_x^2)G_0-2 t_x^2t_nG_0^2}{1-t_pt_nG_0^2+t_x^2t_nG_0^3}
\label{Eq:kamalov}
\end{equation} 
where $t_x=t_{ex}/ \sqrt{1+t_{0}^{(n)} G_0}$.

\section{ $\bar{K} N N$ absorption}

\begin{figure}[h]
\centering
\includegraphics[width=0.8\textwidth] {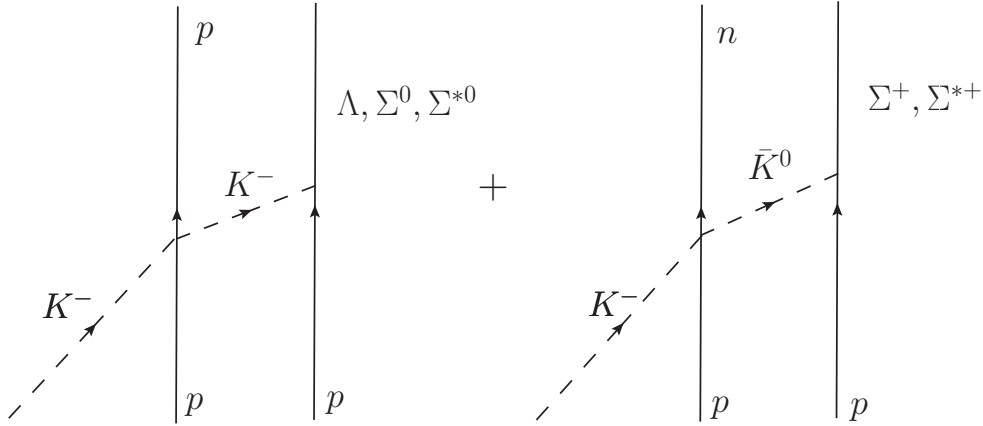}
\caption{Feynman Diagrams for $K^-$ absorption by a proton pair.}
\label{fig:newfig4}
\end{figure}

	We are going to formulate the $\bar{K}$ absorption  by two nucleons. Take for instance $K^{-}$ absorption by a pair of nucleons. The corresponding diagrams for this process are given by Fig \ref{fig:newfig4}
	
 The S-matrix elements for these diagrams are given as follows:

 \begin{eqnarray}
S&=&\frac{1}{V^2} \int \frac{d^{3}q}{(2\pi)^{3}} \frac{1}{\sqrt{2\omega_{\bar{K}}}} t_{\bar{K}N\rightarrow \bar{K}N}  
\frac{1}{q^2-m_{\bar{K}}^2+i \epsilon}~\nonumber \\
&&\times V_{y_{Z(H)}}A(B)~ \tilde \varphi(\vec q- \vec p_{Z(H)}+\frac{\vec P}{2})(2\pi)^{4}
\delta^4(p_{i}-p_{f}) \label{ss11}\\
&\equiv& -iT \frac{1}{\sqrt{2\omega_{\bar{K}}}} \frac{1}{V^2} (2\pi)^{4}\delta^4(p_{i}-p_{f}), \nonumber
\end{eqnarray}	
where $ V_{y_{Z(H)}}A(B) $ is the meson baryon Yukawa vertex, $  Z $ stands for the $ \Lambda, \Sigma^{0},  \Sigma^{+} $ spin-$ \frac{1}{2} $ octet baryons, $  H$ represents the  $ \Sigma^{*0}, \Sigma^{*+}$  spin-$ \frac{3}{2} $  decuplet baryons, $ A$ is equal to $ \vec \sigma \vec q $ for the $\bar{K}N Z  $ vertex and $ B$ is equal to $ \vec S^{\dagger} \vec q $ for the  $\bar{K}N H  $ vertex, where $ \vec{S}^{\dagger} $ is the spin transition operator from spin-$ \frac{1}{2} $ to spin-$ \frac{3}{2} $. Here $ \tilde \varphi(\vec q)$ is the wave function in momentum space of the pair of initials nucleons and $p_i$ and $p_f$ are the initial and final momenta, respectively.
The momentum $\vec p_{Z(H)}$ stands for that of the produced hyperon.

Taking the initial  $ \bar{K}NN $ system at rest, we write the $\bar{K}$ propagator as below

\begin{align}
\frac{1}{q^2-m_{\bar{K}}^2}\rightarrow \frac{1}{(q^{0})^{2}-\vec p^{~2}_{Z(H)}-m_{\bar{K}}^2},
\label{denk10}
\end{align}
where $ q_{Z(H)}^{0}=E_{Z(H)}-E_{N} $ with $E_{Z(H)}=(M_{\bar{K}NN}^{2}+ M_{Z(H)}^{2}-M_{N}^{2})/2M_{\bar{K}NN} $, with $ M_{\bar{K}NN} $ the mass of the $ \bar{K}NN $ system and $ E_N=M_N-\frac{1}{3}B$, where we have subtracted 1/3 of binding energy, $B=m_{\bar{K}}+2M_N-\sqrt{s}$ ($\sqrt{s}=M_{\bar{K}NN}$), to this single nucleon. Since the values of $p_Z$ are around 550 MeV/c, neglecting the Fermi motion of the initial nucleons induces errors of the order of $p_p^2/(p_Z^2+m_{\bar{K}}^2)$ in Eq. (\ref {denk10}), which is about 1 $\%$. 

After redefinition of $\vec q~' \equiv \vec q - \vec p_{Z(H)}$,  the square of the total matrix element summed over the spins of the $ Z $, or $ H $, and averaged over the spin of the nucleons is obtained as

\begin{eqnarray}
 |T|^2 &=&V_{y_{Z(H)}}^2 C_{Z(H)} \vec p^{~2}_{Z(H)} 
 \left(\frac{1}{(q^{0})^{2}-p_{Z(H)}^2-m_{\bar{K}}^2}\right)^2 \nonumber \\
&&\times 
\left|\frac{1}{2 \pi^2}\int q'^2 dq'\tilde \varphi(\vec q~')  t_{{\bar{K}}N,{\bar{K}}N}(\sqrt{s'}) \right|^2
\label{Eq.Tsqeski1}.
 \end{eqnarray}
 where 
$ s'=(p_{\bar{K}}+p_N)^2=m_{\bar{K}}^2+M_N^2+\frac{1}{2} (M_{\bar{K}NN}^{2}-m_{\bar{K}}^2-4M_N^2) $ \cite{multirho,Bayar:2011qj},  with $ C_{Z}=1 $ and $C_{H}=2/3 $.
The formula for $s'$ is obtained assuming that $p_{\bar{K}}.p_N$ is the same for the two nucleons, such that $(p_{\bar{K}} +2 p_N)^2=s$. This neglects recoil effects on the nucleon or the kaons. Effects from this corrections were estimated in \cite{bayarnpa} and affected mildly the binding energy but not the width, and we assume  it to be the case for the absorption width too.

\begin{figure}[h]
\centering
\includegraphics[width=0.8\textwidth] {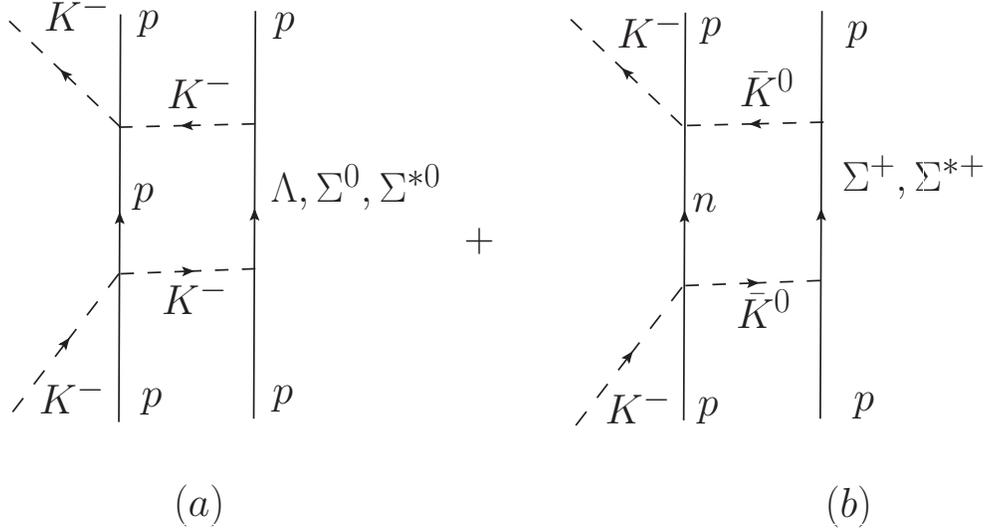}
\caption{Diagrammatic representation of the the $K^- (p p) $ absorption.}
\label{fig:abs1}
\end{figure}

Using the coefficients in Table I and Eq. (14) in Ref. \cite{Oset:2000eg}, we have for the coefficients of the Yukawa vertex 

\begin{eqnarray}
V_{y_{Z}}&=&-\frac{1}{\sqrt{3}}\dfrac{3F+D}{2f} ~~\text{for}~~ K^{-}p \rightarrow \Lambda \nonumber\\
&=&\dfrac{D-F}{2f} ~~\text{for}~~ K^{-}p \rightarrow \Sigma^{0} \nonumber\\
&=&\sqrt{2}\dfrac{D-F}{2f} ~~\text{for}~~ \bar{K}^{0}p \rightarrow \Sigma^{+}
\label{yukavacoeff1}
\end{eqnarray}
with $ D=0.795 $ and $ F=0.465 $. Similarly, for the $\bar{K} N H$ vertex we have 

 \begin{eqnarray}
V_{\bar{K} N H}=a\dfrac{g_{H}}{2M_N}
\end{eqnarray}
 The coefficient $a$ is given in Table 2 in Ref. \cite{Oset:2000eg} and the coupling $ \dfrac{g_{H}}{2M_N} $ is given by, 
\begin{eqnarray}
\dfrac{g_{H}}{2M_N}=\dfrac{2\sqrt{6}}{5}\dfrac{D+F}{2f}.
\end{eqnarray}
Thus, the coefficients for the $ \bar{K} N H $  Yukawa vertices are 
\begin{eqnarray}
V_{y_{H}}&=&\frac{2}{5}\dfrac{F+D}{2f} ~~\text{for}~~ K^{-}p \rightarrow \Sigma^{*0} \nonumber\\
&=&\dfrac{2\sqrt{2}}{5}\dfrac{D+F}{2f} ~~\text{for}~~ \bar{K}^{0}p \rightarrow \Sigma^{*+}
\label{yukavacoeff2}
\end{eqnarray} 
It is useful to relate the  $ T $ matrix with the cross section. Using  $ T $ as the matrix for the transition from $\bar{K}$ and the $(NN)$ cluster to nucleon hyperon $(NZ(H))$, we calculate the $\bar{K} (NN)\rightarrow N Z(H)$ cross section as 
 \begin{eqnarray}
 \sigma_{\text{abs}} = \frac{1}{2 \pi} \frac{ M_{NN}  M_{Z(H)} M_N }{M_{\bar{K}NN}^2 }  \frac{p_{Z(H)}}{p_{\bar{K}}}|T|^2.
 \nonumber
\end{eqnarray}
It is interesting to relate this cross section with the imaginary part of the $\bar{K}(NN) \rightarrow \bar{K}(NN)$ amplitude for the diagrams of Fig. \ref{fig:abs1}.
Borrowing the optical theorem we rewrite the cross section in terms of the imaginary part of the $\bar{K}(NN) \rightarrow \bar{K}(NN)$ amplitude, $ T_{\bar{K} (NN)} $, as follow: 

\begin{eqnarray}
\text{Im }T_{\bar{K}(NN)}=-\frac{p_{\bar{K}} \sqrt{s}}{M_{NN}} \sigma_{\text{abs}} = -\frac{1}{2 \pi} \frac{M_{Z(H)} M_N }{M_{\bar{K}NN}} p_{Z(H)} |T|^2.\label{Eq:imtkbarnn1}
\end{eqnarray} 

\begin{figure}
\centering
\includegraphics[width=0.8\textwidth] {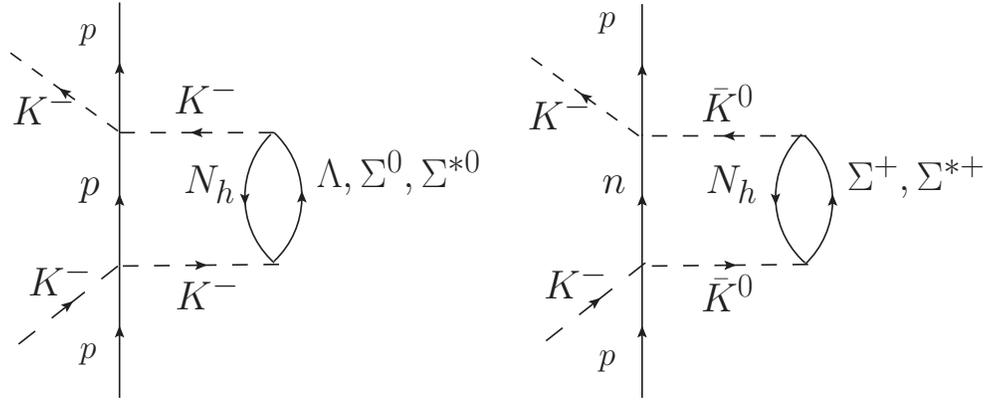}
\caption{Representation of the $K^- (p p) $ absorption in many-body diagrams.}
\label{fig:abs2}
\end{figure}

 In analogy to \cite{bayardnn}, we can convert the absorption diagram of  Fig. \ref{fig:abs1} into the "many-body" diagram of Fig. \ref{fig:abs2}, where the nucleon on which the virtual $ \bar{K} $ is absorbed is converted into a "hole" line. The purpose of following this path is that we can now include the absorption process by making a modification of the meson baryon loop function, $ \delta G $, to include the "ph" excitation in the $ \bar{K} $ propagator. Then, we can  reevaluate the $ \bar{K} N $ amplitude in the coupled channels nonperturbative Bethe Salpeter equations and use the new amplitude in the $\bar{K} (NN)$ amplitudes of Eqs. (\ref{Eq:bizim}) and (\ref{Eq:kamalov}) which sum the multiple scattering series of the $\bar{K} (NN)$ system. In this way we perform fully nonperturbatively the implementation of the $\bar{K} (NN)$ absorption in the $\bar{K} (NN)$ system.
 
 Since we are only concerned about the absorption process it suffices to evaluate $ \text{Im } \delta G $, but later on we shall also look at the real part. This is easy since  $ \text{Im } \delta G $ is the same as $ \text{Im } T_{\bar{K} (NN)} $ removing the modulus squared of the  $ t_{\bar{K}N,\bar{K}N} $ amplitude in Eq. (\ref{Eq.Tsqeski1}). This amplitude, needed for the physical process, is regained when we evaluate the $\bar{K}N$ amplitudes in the chiral unitary approach substituting $G_{\bar{K}N}$ by $G_{\bar{K}N}+ \delta G_{\bar{K}N}$ (See Ref. \cite{bayardnn} for details).
 Thus we can immediately write
 
\begin{eqnarray}
i \text{Im } \delta G_{K^-p} = -i\frac{1}{2 \pi} \frac{ M_N }{M_{\bar{K}NN}} \sum_{i=1}^{3} M_{Z(H)} p_{Z(H)} | T_{Z(H)}|^2, \label{Eq:deltaG1}
\end{eqnarray}

\begin{eqnarray}
i \text{Im } \delta G_{\bar{K}^0 n} = -i\frac{1}{2 \pi} \frac{ M_N }{M_{\bar{K}NN}} \sum_{i=1}^{2} M_{Z(H)} p_{Z(H)} | T_{Z(H)}|^2 \label{Eq:deltaG2}
\end{eqnarray}
where $T_{Z(H)}$ is the $T$ matrix defined before, specifying the final hyperon, removing the modulus squared of $ t_{\bar{K}N,\bar{K}N} $.
The sum for 1-3 of $  \delta G_{K^-p} $  runs over $ \Lambda, ~\Sigma^{0} $ for type $ Z $ and $ \Sigma^{*0} $ for type $ H $, and the sum 1-2 of $  G_{\bar{K}^0 n} $ runs over $ \Sigma^{+} $ of type $ Z $ and $\Sigma^{*+}  $ of type $ H $ (see Fig. \ref{fig:abs2}).

 In Eqs. (\ref{Eq:deltaG1}) and (\ref{Eq:deltaG2}), $  |T_{Z(H)}|^2 $ is given by

\begin{eqnarray}
 |T_{Z(H)}|^2 &=&|\varphi (0)|^2 V_{y_{Z(H)}}^2 C_{Z(H)}  \vec p^{~2}_{Z(H)} 
 \left(\frac{1}{(q_i^{0})^{2}-p_{Z(H)}^2-m_{\bar{K}}^2}\right)^2 
\label{Eq.ttilda}
 \end{eqnarray}
where $ V_{y_{Z}} $, $ V_{y_{H}} $, are given in Eqs. (\ref{yukavacoeff1}) and (\ref{yukavacoeff2}), respectively. Note that when removing the $ t_{\bar{K} N,\bar{K} N} $ amplitude in the integral of Eq. (\ref{Eq.Tsqeski1}), the remaining integral simply gives the wave function in coordinate space in the origin, $ \varphi (0) $.

For $ \varphi (r) $ we take the same form of wave function as in \cite{bayardnn} for the $ NN $ cluster

\begin{eqnarray}
\varphi(r)=a e^{-\alpha r},~
\quad a = \frac{1}{2} 
\left(\frac{\alpha^3}{2\pi}\right)^{\frac{1}{2}},
\tilde{\varphi} (q) = \frac{4 \pi a \alpha}{(\frac{1}{4} \alpha^2 - q^2)^2 + q^2\alpha^2}
\end{eqnarray}
but with parameters suited to the $NN$ wave function in the $\bar{K} NN$ cluster. For this,
the parameter $ \alpha$ is the chosen such as to get the $NN$ cluster with a relative distance of around 2 fm as found in \cite{Dote:2008in,Dote:2008hw} for the present problem. We accomplish this with  $ \alpha=1.7 ~$ fm$^{-1}$.
The momentum of the baryons are given by 
\begin{eqnarray}
p_{Z(H)}=\dfrac{\lambda^{\frac{1}{2}}(M_{\bar{K}NN}^{2},M_{N}^{2},M_{Z(H)}^{2})}{2M_{\bar{K}NN}}
\label{pzh}
\end{eqnarray}
and in order to account for relativistic corrections, $\vec p^{~2}_{Z(H)} $ accompanied by $  V_{y_{Z(H)}}^2 $ is given by 
\begin{eqnarray}
\vec p^{~2}_{Z(H)}V_{y_{Z(H)}}^2 \longrightarrow V_{y_{Z(H)}}^2 \dfrac{1}{4 M_{Z(H)}^{2}}(M_N+ M_{Z(H)})^{2} \vec p^{~2}_{Z(H)}
\end{eqnarray}

Since we get different contributions for $ \delta G $ in the $ K^{-} p$ or $ \bar{K}^{0}n $ intermediate states, we are now forced to recalculate the $ \bar{K}N $ amplitudes using the charge basis of the coupled channels, rather than the isospin base normally used.

\begin{figure}[h]
\centering
\includegraphics[width=0.7\textwidth] {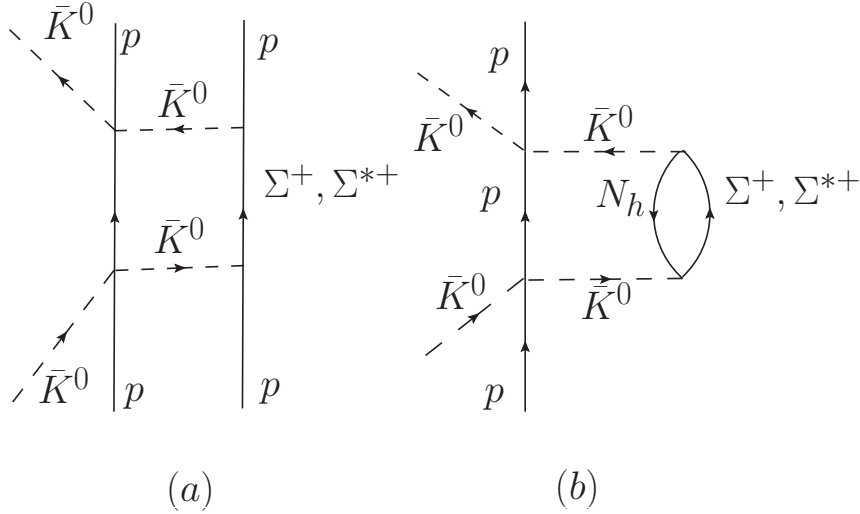}
\caption{Representation of the $K^- (p p) $ absorption for $ t_{\bar{K}^{0}p \rightarrow \bar{K}^{0}p} $ amplitude.}
\label{fig:abs3}
\end{figure}

Furthermore, as seen in Eqs. (\ref{Eq:partf}), we also  need to calculate the $ t_{\bar{K}^{0}p \rightarrow \bar{K}^{0}p} $ amplitude. This is a channel that has only one Feynman diagram as shown in Fig.  \ref{fig:abs3}. In order to calculate the $ t_{\bar{K}^{0}p \rightarrow \bar{K}^{0}p} $ amplitude we use the same potential with $ V_{K^{-}n,K^{-}n} $ \cite{Oset:1997it}. The $ \bar{K}^{0}p $ channel is renormalized in the same way as the  $\bar{K}^{0}n  $, as one can see, comparing Fig. \ref{fig:abs3} (a) with Fig. \ref{fig:abs1} (b). Hence, one can use  $ G_{\bar{K}^{0}p} \rightarrow G_{\bar{K}^{0}p} +i \text{Im }\delta G_{\bar{K}^0 n} $ for the channel $ \bar{K}^{0}p $.

We have been concerned about the consideration of the absorption channel in the width of the $ \bar{K}NN $ system. With a bit more of effort we can equally calculate the effects of the real part associated to the absorption (dispersive part). For this it is sufficient to substitute $i \text{Im } G_{\bar{K} N} $ of Eqs. (\ref{Eq:deltaG1}), (\ref{Eq:deltaG2})  by $\delta G$ given by 

\begin{eqnarray}
\delta G&=&i\int \frac{d^4p}{(2\pi)^4} \frac{M_N M_{Z(H)}}{E_{N}(\vec{p}) E_{Z(H)}(\vec{p})} |\varphi (0)|^2 V_{y_{Z(H)}}^2 C_{Z(H)}  \vec p^{~2}   \frac{1}{p^0-E_{N}(\vec{p})+i \epsilon}\nonumber\\&&
\Big[\frac{1}{(E_{\bar{K}N}(\vec{p})-p^0)^2-\omega(\vec{p})^2+i \epsilon}\Big]^2
\frac{1}{E_{\bar{K}N}(\vec{p})-p^0+p_2^0-E_{Z(H)}(\vec{p})+i \epsilon} 
\label{deltaGfourmomentum}
 \end{eqnarray}
where $E_{\bar{K}N}$ is the energy of the initial $K^-$ and the first nucleon of the $\bar{K} NN$ system and $p_2^0$ is the energy of the second nucleon. We calculate them splitting equally the binding energy B between the three hadrons then
  
 \begin{eqnarray} 
E_{\bar{K}N}&=&m_{\bar{K}}+M_N-\frac{2}{3}B
\nonumber\\
p_2^0&=&M_N-\frac{1}{3}B.
 \end{eqnarray}
The $p^0$ integration in Eq.(\ref{deltaGfourmomentum}) can be performed analytically and we find 

\begin{eqnarray}
\delta G&=&\int \frac{d^3p}{(2\pi)^3} \frac{M_N M_{Z(H)}}{E_{N}(\vec{p}) E_{Z(H)}(\vec{p})} 
|\varphi (0)|^2 V_{y_{Z(H)}}^2 C_{Z(H)}  \vec p^{~2} \Big(\frac{1}{2\omega(\vec{p})}\Big)^2
\nonumber\\&&(\sqrt{s}-E_{N}(\vec{p})-2\omega(\vec{p})-E_{Z(H)}(\vec{p}))
\frac{1}{E_{\bar{K}N}(\vec{p})-E_{N}(\vec{p})-\omega(\vec{p})+i \epsilon}\nonumber\\&&
\frac{1}{\sqrt{s}-E_{N}(\vec{p})-E_{Z(H)}(\vec{p})+i \epsilon}~~
\frac{1}{-\omega(\vec{p})+p_2^0-E_{Z(H)}+i \epsilon} 
\nonumber\\&&\Big[\frac{-2}{\sqrt{s}-E_{N}(\vec{p})-2\omega(\vec{p})-E_{Z(H)}(\vec{p})}-\frac{1}{\omega(\vec{p})}\nonumber\\&&+\frac{1}{E_{\bar{K}N}(\vec{p})-E_{N}(\vec{p})-\omega(\vec{p})+i \epsilon}+\frac{1}{-\omega(\vec{p})+p_2^0-E_{Z(H)}+i \epsilon}\Big]
\label{deltaGthreemomentum}
 \end{eqnarray}
where $E_{Z(H)}(\vec{p})=\sqrt{M_{Z(H)}^2+\vec{p}^{~2}}$, $\omega(\vec{p})=\sqrt{m_{\bar{K}}^2+\vec{p}^{~2}}$, $E_{N}(\vec{p})=\sqrt{M_{N}^2+\vec{p}^{~2}}$, from where one can easily see that $ \text{Im }\delta G  $ corresponds to what we found in Eqs. (\ref{Eq:deltaG1}) and (\ref{Eq:deltaG2}). We use a $q_{max}$ in the integration of $ \delta G  $, Eq. (\ref{deltaGthreemomentum}), of $q_{max}=630$ MeV as done in \cite{Oset:1997it}.
 
  We are also interested in learning the effects of making the nonperturbative approach that we have done compared to one typical perturbative approach. We can make use of the same formalism that we have followed using an easy algorithm: The perturbative calculation contains only one bubble of nucleon hole-$Z(H)$. We can regain the one bubble results by substituting $|\phi(0)|^2 \rightarrow |\phi(0)|^2 \beta$ with $\beta$ a parameter running from zero to 1. We then take the tangent of the curve in $\beta$ at the origin of $\beta$. The result of this linear extrapolation at $\beta=1$ gives us the perturbative results for the amplitude.

\section{ Results}

As we stated in the Introduction, the $ \Lambda(1405)$ plays a key role in the $ \bar{K} $ absorption. As mentioned before, we recalculate the $ \bar{K} N $ amplitudes in the charge base.
In the case of the $Q=0$ and $I_3=0$, there are ten coupled channels which are $ K^{-}p,~\bar{K}^0 n,~ \pi^{0}\Lambda,~\pi^{0} \Sigma^{0},~ \eta\Lambda, ~\eta \Sigma^{0},~\pi^{+}\Sigma^{-},~\pi^{-}\Sigma^{+},~ K^{+} \Xi^{-},~ K^{0} \Xi^{0}$ \cite{Oset:1997it}. As discussed above, we also need the $\bar{K}^{0}p \rightarrow  \bar{K}^{0}p  $ amplitude in pure $ I=1 $. For this we take the  $ I_{3}=-1 $ component and use the 
 six coupled channels, $K^{-}n,~ \pi^{0} \Sigma^{-},~ \pi^{-} \Sigma^{0},~ \pi^{-} \Lambda,~ \eta\Sigma^{-},~ K^{0} \Xi^{-} $ \cite{Oset:1997it}. In the coupled channels reevaluation of the $ \bar{K} N$ amplitudes for $ I_{3}=0 $ we add $ i~\text{Im }\delta G_{K^-p} $, $ i~\text{Im }\delta G_{\bar{K}^0 n} $ to $ G_{K^-p} $, $ G_{\bar{K}^0 n} $, respectively. For $ I_{3}=1 $ we add $ i~\text{Im } \delta G_{K^-n} $ to $ G_{K^-n} $, but for isospin symmetry reasons $ i~\text{Im }\delta G_{K^-n} \equiv i ~\text{Im }\delta G_{\bar{K}^0 p} $ which we have discussed above. In all the channels we use the same numeric value for $ f $, $f=1.123 f_{\pi}$ ($ f_{\pi} $=93 MeV) as in \cite{Oset:1997it}. For the form factor of Eq. (\ref{Eq:gzero}) we use the deuteron type form factor with the reduced size of the two N system found in \cite{Dote:2008hw}.

As discussed in the former section we also perform the calculations using the full $\delta G$, that contains the real part associated to the absorption mechanism. In addition we also evaluate the results in the perturbative case for the absorption mechanism. 
 
\begin{figure}[b]
\centering
\includegraphics[width=0.9\textwidth] {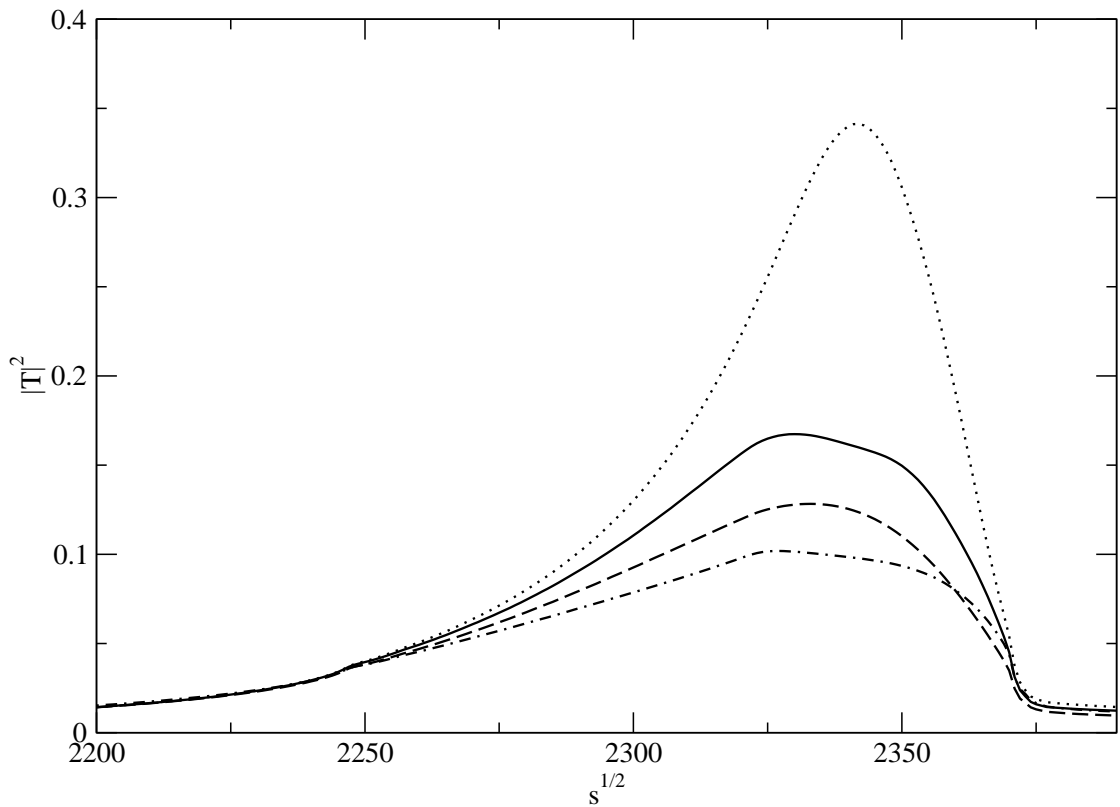}
\caption{Modulus squared of the $ T $ matrix for $ \bar{K} $ scattering on the $ NN $ cluster for $ S=0 $. The doted line indicates the result without absorption. The dashed line indicates the result including the contribution of $ \bar{K} $ absorption on two nucleons. The dash-doted line indicates the effect of the real part and the solid line indicates the result obtained in the nonperturbative calculation using only $\text{Im } \delta G$.}
\label{fig:TsqrS0}
\end{figure}

\begin{figure}
\centering
\includegraphics[width=0.9\textwidth] {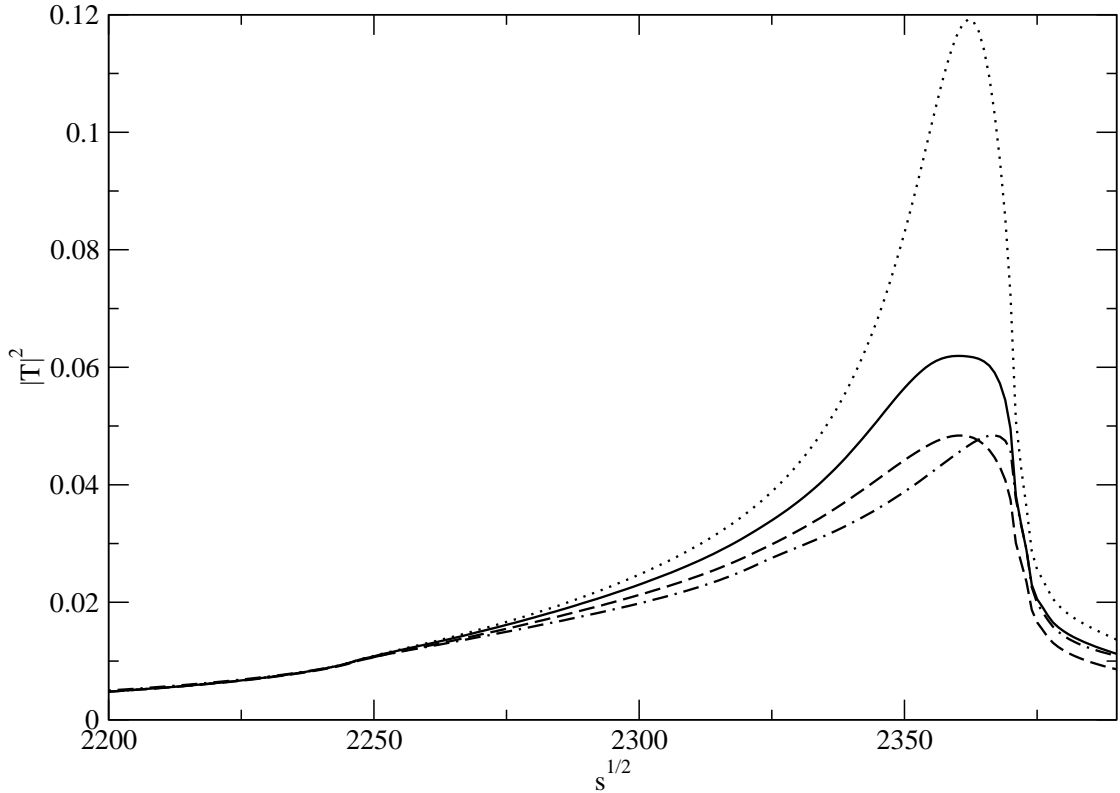}
\caption{Same as Fig. \ref{fig:TsqrS0} for $ S=1 $.}
\label{fig:TsqrS1}
\end{figure}

In Figs. \ref{fig:TsqrS0} and  \ref{fig:TsqrS1} we show the absolute squared of the $ T $ matrix
for $ \bar{K} $ scattering on the $NN $ cluster $ S=0 $, including absorption,  for the cases of $ S=0 $ and $ S=1 $. As we can see from Fig. \ref{fig:TsqrS0}, the most striking thing is a substantial increase in the width, that goes from about 50 MeV to about 80 MeV, and which is due to absorption.  The centroid of the distribution is also a bit displaced to lower energies, but what concerns us now is that the  $ \bar{K} $ absorption on two nucleons has increased the width by about 30 MeV. The order of magnitude is similar to what was estimated in \cite{Dote:2008in,Dote:2008hw}, but here we have done a full nonperturbative evaluation of this magnitude. In the case of $ S=1 $ in Fig. \ref{fig:TsqrS1} we observe that the shape of the distribution has been distorted considerably due to the consideration of the $ \bar{K} $ absorption and the centroid has not changed appreciably. From this distorted shape one  can also estimate that the width has increased in about 30 MeV, like in the previous case.  The width of the state is estimated from the width of the curve at half the value of its maximum, even if the shape is not that of a Breit Wigner distribution.

We should mention that we have also included the effect  of the $ \pi \Sigma N $ intermediate states in the calculation. Here also the  $\bar{K} N \rightarrow \pi \Sigma $ amplitudes have been modified due to absorption, since they come from the coupled channels calculation. The contribution of this channel has been evaluated as in \cite{Bayar:2011qj} and its effect is small, like also found there.

In the figures we can see the effect of considering the real part of $\delta G$. We observe that the inclusion of the real part does not practically modify the results. 
One appreciates a small increase in the width by about 5 MeV for the case of $S=0$. We also show in the figures the result obtained in the nonperturbative calculation. This has been done as discussed in the former section but just in the case of taking only $\text{Im } \delta G$,
 which we found to be the relevant part. For $S=0$, we can see that, with respect to the nonperturbative calculation the strength at the peak in $|T|^2$ is reduced by about 20$\%$ and the width is increased by about 5 MeV.
 
 The results for $S=1$ are similar, the only difference being that when the full $\delta G$ is considered the width decreases by about 5 MeV in this case and a small shift on the peak to bigger energies by the same amount is produced.

\newpage
\section{ Conclusions}

We have made a detailed calculation of the contribution to the width of the bound  $ \bar{K} NN $ states from $ \bar{K} $ absorption on two nucleons. The evaluation is done nonperturbatively in two aspects: First, the $ \bar{K} N $ amplitudes are reevaluated  in the unitary coupled channels approach taking into account the absorption of the $ \bar{K} $. Second the resulting $ \bar{K} N $ amplitudes are used in the nonperturbative formula of the Fixed Center Approximation that takes into account the rescattering of the kaons on the nucleons of the $ NN $ cluster.

 The result of these calculations is that the width of the states with $ S=0 $, $ S=1 $ is increased by about 30 MeV to values of the total width of 75-80 MeV. The relative simplicity of the formalism has allowed us to address some issues for the first time, as the nonperturbative evaluation of the absorption width and the effects of the real (dispersive) part associated to absorption. 
 
  With the large width obtained and the small values of the binding, 15-30 MeV, to which the different groups are converging \cite{Dote:2008hw,Ikeda:2010tk,Barnea:2012qa}, we are facing a situation of states with much larger width than binding, which makes the experimental  observation problematic. Further calculations taking advantage of the steps given in the present paper, but using different formalisms, would be most welcome.

  \section{Acknowledgments}
  We thank A. Gal and T. Hyodo for useful comments. This work is partly supported by DGICYT contract number
FIS2011-28853-C02-01, the Generalitat Valenciana in the program Prometeo, 2009/090. We acknowledge the support of the European Community-Research Infrastructure
Integrating Activity
Study of Strongly Interacting Matter (acronym HadronPhysics3, Grant Agreement
n. 283286)
under the Seventh Framework Programme of EU.



\begin{thebibliography}{999}

\bibitem{Kaiser:1995eg}
  N.~Kaiser, P.~B.~Siegel and W.~Weise,
  Nucl.\ Phys.\  A {\bf 594}, 325 (1995).

\bibitem{angels}
  E.~Oset and A.~Ramos,
  Nucl.\ Phys.\  A {\bf 635}, 99 (1998).

\bibitem{Oller:2000fj}
  J.~A.~Oller and U.~G.~Meissner,
  Phys.\ Lett.\  B {\bf 500}, 263 (2001).

\bibitem{Lutz:2001yb}
  M.~F.~M.~Lutz and E.~E.~Kolomeitsev,
  Nucl.\ Phys.\  A {\bf 700}, 193 (2002).
  
\bibitem{Oset:2001cn}
  E.~Oset, A.~Ramos and C.~Bennhold,
  Phys.\ Lett.\  B {\bf 527}, 99 (2002)
  [Erratum-ibid.\  B {\bf 530}, 260 (2002)].
 
\bibitem{Hyodo:2002pk}
  T.~Hyodo, S.~I.~Nam, D.~Jido and A.~Hosaka,
  Phys.\ Rev.\  C {\bf 68}, 018201 (2003).

\bibitem{Jido:2003cb}
  D.~Jido, J.~A.~Oller, E.~Oset, A.~Ramos and U.~G.~Meissner,
  Nucl.\ Phys.\  A {\bf 725}, 181 (2003).


\bibitem{Borasoy:2005ie}
  B.~Borasoy, R.~Nissler and W.~Weise,
  Eur.\ Phys.\ J.\  A {\bf 25}, 79 (2005).

\bibitem{Oller:2006jw} 
  J.~A.~Oller,
  Eur.\ Phys.\ J.\ A {\bf 28}, 63 (2006)
  [hep-ph/0603134].
  
\bibitem{Borasoy:2006sr} 
  B.~Borasoy, U.~-G.~Meissner and R.~Nissler,
  Phys.\ Rev.\ C {\bf 74}, 055201 (2006)
  [hep-ph/0606108].
 
\bibitem{Ikeda:2012au} 
  Y.~Ikeda, T.~Hyodo and W.~Weise,
  Nucl.\ Phys.\ A {\bf 881}, 98 (2012)
  [arXiv:1201.6549 [nucl-th]].

 

\bibitem{nogami} Y. Nogami, Phys. Lett. 7, 288 (1963).

 


\bibitem{Yamazaki:2002uh}
  T.~Yamazaki and Y.~Akaishi,
  Phys.\ Lett.\  B {\bf 535} (2002) 70.
  
 

  
\bibitem{Dote:2008in}
  A.~Dote, T.~Hyodo and W.~Weise,
  Nucl.\ Phys.\  A {\bf 804}, 197 (2008).


\bibitem{Dote:2008hw}
  A.~Dote, T.~Hyodo and W.~Weise,
  Phys.\ Rev.\  C {\bf 79}, 014003 (2009)
  [arXiv:0806.4917 [nucl-th]].


\bibitem{Shevchenko:2006xy}
  N.~V.~Shevchenko, A.~Gal and J.~Mares,
  Phys.\ Rev.\ Lett.\  {\bf 98}, 082301 (2007)
  [arXiv:nucl-th/0610022].
  
\bibitem{Shevchenko:2007zz}
  N.~V.~Shevchenko, A.~Gal, J.~Mares and J.~Revai,
  Phys.\ Rev.\  C {\bf 76}, 044004 (2007)

\bibitem{Ikeda:2007nz} 
  Y.~Ikeda and T.~Sato,
  Phys.\ Rev.\ C {\bf 76}, 035203 (2007)
  [arXiv:0704.1978 [nucl-th]].

\bibitem{Ikeda:2008ub} 
  Y.~Ikeda and T.~Sato,
  Phys.\ Rev.\ C {\bf 79}, 035201 (2009)
  [arXiv:0809.1285 [nucl-th]].
  
\bibitem{Ikeda:2010tk} 
  Y.~Ikeda, H.~Kamano and T.~Sato,
  Prog.\ Theor.\ Phys.\  {\bf 124}, 533 (2010)
  [arXiv:1004.4877 [nucl-th]].



\bibitem{bayarnpa} 
  M.~Bayar and E.~Oset,
  Nucl.\ Phys.\ A {\bf 883}, 57 (2012)
  [arXiv:1203.5313 [nucl-th]].
  

 
\bibitem{Oset:2012gi} 
  E.~Oset, D.~Jido, T.~Sekihara, A.~M.~Torres, K.~P.~Khemchandani, M.~Bayar and J.~Yamagata-Sekihara,
  Nucl.\ Phys.\ A {\bf 881}, 127 (2012)
  [arXiv:1203.4798 [hep-ph]].

\bibitem{Barnea:2012qa} 
  N.~Barnea, A.~Gal and E.~Z.~Liverts,
  Phys.\ Lett.\ B {\bf 712}, 132 (2012)
  [arXiv:1203.5234 [nucl-th]].


\bibitem{Xie:2010ig} 
  J.~-J.~Xie, A.~Martinez Torres and E.~Oset,
  Phys.\ Rev.\ C {\bf 83}, 065207 (2011)
  [arXiv:1010.6164 [nucl-th]].

\bibitem{Jido:2008kp} 
  D.~Jido and Y.~Kanada-En'yo,
  Phys.\ Rev.\ C {\bf 78}, 035203 (2008)
  [arXiv:0806.3601 [nucl-th]].

\bibitem{MartinezTorres:2008kh} 
  A.~Martinez Torres, K.~P.~Khemchandani and E.~Oset,
  Phys.\ Rev.\ C {\bf 79}, 065207 (2009)
  [arXiv:0812.2235 [nucl-th]].


\bibitem{Xie:2011uw} 
  J.~-J.~Xie, A.~Martinez Torres, E.~Oset and P.~Gonzalez,
  Phys.\ Rev.\ C {\bf 83}, 055204 (2011)
  [arXiv:1101.1722 [nucl-th]].

\bibitem{MartinezTorres:2010ax} 
  A.~Martinez Torres, E.~J.~Garzon, E.~Oset and L.~R.~Dai,
  Phys.\ Rev.\ D {\bf 83}, 116002 (2011)
  [arXiv:1012.2708 [hep-ph]].

\bibitem{Friedman:1999rh} 
  E.~Friedman and A.~Gal,
  Phys.\ Lett.\ B {\bf 459}, 43 (1999)
  [nucl-th/9902036].



\bibitem{Koch:1994mj} 
  V.~Koch,
  Phys.\ Lett.\ B {\bf 337}, 7 (1994)
  [nucl-th/9406030].

\bibitem{Lutz:1997wt} 
  M.~Lutz,
  Phys.\ Lett.\ B {\bf 426}, 12 (1998)
  [nucl-th/9709073].

\bibitem{Ramos:1999ku} 
  A.~Ramos and E.~Oset,
  Nucl.\ Phys.\ A {\bf 671}, 481 (2000)
  [nucl-th/9906016].

\bibitem{SchaffnerBielich:1999cp} 
  J.~Schaffner-Bielich, V.~Koch and M.~Effenberger,
  Nucl.\ Phys.\ A {\bf 669}, 153 (2000)
  [nucl-th/9907095].
  
\bibitem{Cieply:2001yg} 
  A.~Cieply, E.~Friedman, A.~Gal and J.~Mares,
  Nucl.\ Phys.\ A {\bf 696}, 173 (2001)
  [nucl-th/0104087].

\bibitem{Ramos:2008zza}  A.~Ramos, V.~K.~Magas, E.~Oset and H.~Toki,
 Nucl.\ Phys.\ A {\bf 804}, 219 (2008).

\bibitem{Hirenzaki:2000da}  S.~Hirenzaki, Y.~Okumura, H.~Toki, E.~Oset and A.~Ramos,
 Phys.\ Rev.\ C {\bf 61}, 055205 (2000).
  
  

\bibitem{Veirs:1970fs} 
  V.~R.~Veirs and R.~A.~Burnstein,
  Phys.\ Rev.\ D {\bf 1}, 1883 (1970).

\bibitem{Katz:1970ng} 
  P.~A.~Katz, K.~Bunnell, M.~Derrick, T.~Fields, L.~G.~Hyman and G.~Keyes,
  Phys.\ Rev.\ D {\bf 1}, 1267 (1970).
  
\bibitem{Oset:2000eg} 
  E.~Oset and A.~Ramos,
  Nucl.\ Phys.\ A {\bf 679}, 616 (2001)
  [nucl-th/0005046].
   
\bibitem{Sekihara:2012wj} 
  T.~Sekihara, J.~Yamagata-Sekihara, D.~Jido and Y.~Kanada-En'yo,
  arXiv:1204.3978 [nucl-th].

\bibitem{Okada:2007ky} 
  S.~Okada, G.~Beer, H.~Bhang, M.~Cargnelli, J.~Chiba, S.~Choi, C.~Curceanu and Y.~Fukuda {\it et al.},
  Phys.\ Lett.\ B {\bf 653}, 387 (2007)
  [arXiv:0707.0448 [nucl-ex]].

\bibitem{Bazzi:2009zz} 
  M.~Bazzi {\it et al.}  [SIDDHARTA Collaboration],
  Phys.\ Lett.\ B {\bf 681}, 310 (2009).
 
\bibitem{Baca:2000ic}  A.~Baca, C.~Garcia-Recio and J.~Nieves,
 Nucl.\ Phys.\ A {\bf 673}, 335 (2000)
 [nucl-th/0001060].
  
  
  
  










  


  
 


 
 
\bibitem{multirho}
  L.~Roca and E.~Oset,
  Phys.\ Rev.\  D {\bf 82}, 054013 (2010).

\bibitem{Bayar:2011qj} 
  M.~Bayar, J.~Yamagata-Sekihara and E.~Oset,
  Phys.\ Rev.\ C {\bf 84}, 015209 (2011)
  [arXiv:1102.2854 [hep-ph]].

\bibitem{kamalov}
  S.~S.~Kamalov, E.~Oset and A.~Ramos,
  Nucl.\ Phys.\  A {\bf 690}, 494 (2001).


\bibitem{bayardnn} 
  M.~Bayar, C.~W.~Xiao, T.~Hyodo, A.~Dote, M.~Oka and E.~Oset,
  arXiv:1205.2275 [hep-ph].



\bibitem{Oset:1997it} 
  E.~Oset and A.~Ramos,
  Nucl.\ Phys.\ A {\bf 635}, 99 (1998)




\end{thebibliography}
\end{document}